\documentclass[prl,aps,twocolumn]{revtex4}

\usepackage{psfrag}
\usepackage{graphicx} 
\usepackage{dcolumn}
\usepackage{color} 
\usepackage{latexsym,amsfonts} 
\usepackage{bm}
\usepackage{amssymb} 
\begin{document}

\title{Molecule model for deeply bound and broad kaonic nuclear
clusters}

\author{A. N. Ivanov${^a}$, P. Kienle${^{b,c}}$, J. Marton${^{b}}$,
M. Pitschmann${^{a,d}}$} \affiliation{${^a}$Atominstitut der
\"Osterreichischen Universit\"aten, Technische Universit\"at Wien,
Wiedner Hauptstrasse 8-10, A-1040 Wien, Austria}
\affiliation{${^b}$Stefan Meyer Institut f\"ur subatomare Physik
\"Osterreichische Akademie der Wissenschaften, Boltzmanngasse 3,
A-1090, Wien, Austria}\affiliation{${^c}$Excellence Cluster Universe
Technische Universit\"at M\"unchen, D-85748 Garching,
Germany}\affiliation{$^d$University of Wisconsin--Madison, Department
of Physics, 1150 University Avenue, Madison, WI 53706, USA}
\email{ivanov@kph.tuwien.ac.at}

\date{\today}

\begin{abstract}
 A molecule model is proposed for the description of the properties of
  the kaonic nuclear cluster (KNC) $\bar{K}NN$ with the structure
  $N\,\otimes (\bar{K}N)_{I = 0}$ and quantum numbers $I(J^P) =
  \frac{1}{2}(0^-)$, the large binding energy $B^{(\exp)}_{\bar{K}NN}
  = 103(6)\,{\rm MeV}$ and the width $\Gamma^{(\exp)}_{\bar{K}NN} =
  118(13)\,{\rm MeV}$, observed recently by the DISTO Collaboration.
  The theoretical values of the binding energy $B^{(\rm
  th)}_{\bar{K}NN} = 118\,{\rm MeV}$, the width $\Gamma^{(\rm
  th)}_{\bar{K}NN} = 142\,{\rm MeV}$ of the KNC $\bar{K}NN$ and the
  density $n_{\bar{K}NN} = 2.71\,n_0$, where $n_0= 0.17\,{\rm
  fm^{-3}}$ is the normal nuclear density, reproduce well the large
  experimental values. They are calculated with the trial harmonic
  oscillator wave functions by using chiral Lagrangians, accounting
  for all self--energy terms, contributing to the masses of the kaonic
  nuclear clusters $(\bar{K}N)_{I = 0}$ and $\bar{K}NN$. In addition
  the high $p\Lambda^*$ sticking probability in the $pp$ reaction at
  the kinetic energy $T_p = 2.85\,{\rm GeV}$ of the incident proton is
  explained. 
\end{abstract}  

\maketitle

\subsection{1. Introduction}

Recently \cite{KNC09} we have proposed a molecule model for kaonic
nuclear clusters (KNCs) $\bar{K}N$ and $\bar{K}NN$, which we denote as
${^1_{\bar{K}}}{\rm H}$ and ${^2_{\bar{K}}}{\rm H}$, with the
structure $(\bar{K}N)_{I = 0}$ and $N\otimes (\bar{K}N)_{I = 0}$,
respectively. The binding energies of the KNCs ${^1_{\bar{K}}}{\rm H}$
and ${^2_{\bar{K}}}{\rm H}$ have been calculated in the
tree--approximation and defined by the Weinberg--Tomozawa (WT)
interactions $(\bar{K}N)_{I = 0} \to \bar{K}N)_{I = 0}$ and
$(\bar{K}N)_{I = 1} \to \bar{K}N)_{I = 1}$ only as
$B_{{^2_{\bar{K}}}{\rm H}} = 40.2\,{\rm MeV}$. The WT--interaction
$(\bar{K}N)_{I = 1} \to \bar{K}N)_{I = 1}$ contributed to the binding
energy of the KNC ${^2_{\bar{K}}}{\rm H}$ only due to the $NN$
exchange interaction.

We have shown that in the tree--approximation the molecule model of
the KNCs reproduces well the binding energy $B_{{^2_{\bar{K}}}{\rm H}}
= 48\,{\rm MeV}$ and width of the KNC ${^2_{\bar{K}}}{\rm H}$,
obtained in the potential model by Akaishi and Yamazaki
\cite{Akaishi1}--\cite{Akaishi3}. The binding energy and the width of
the KNC ${^2_{\bar{K}}}{\rm H}$, calculated in the molecule model of
KNCs in the tree--approximation, agree qualitatively also with the
results, obtained in other theoretical approaches
\cite{KNC1}--\cite{KNC3}, but disagree with the $SU(3)$
coupled--channel approach with chiral dynamics, predicting rather
shallow bindings \cite{KNC4,KNC5}.  In this approach two resonances
exist with the same quantum numbers and different masses instead of
the $\Lambda(1405)$ (or the $\Lambda^*$) resonance with strangeness $S
= -1$, quantum numbers $I(J^P) = 0(\frac{1}{2}^-)$ and mass
$m_{\Lambda^*} = 1405\,{\rm MeV}$.  One of these resonances with mass
$m_{\Lambda^*_1} = 1420\,{\rm MeV}$, treated as a quasi-bound
$(\bar{K}N)_{I = 0}$ state, leads to the shallow binding of about
$20\,{\rm MeV}$ of the $\bar{K}NN$ state.  This is unlike the
potential model approach, developed in
\cite{Akaishi1}--\cite{Akaishi3}, and our molecule model, dealing with
the $\Lambda^*$ resonance with mass $m_{\Lambda^*} = 1405\,{\rm MeV}$
only, treated as a quasi-bound $(\bar{K}N)_{I = 0}$ state.

For the calculation of the binding energies and widths of the KNCs
${^1_{\bar{K}}}{\rm H}$ and ${^2_{\bar{K}}}{\rm H}$ we use trial
harmonic oscillator wave functions, parameterised by two frequencies
$\Omega_{\Lambda^*}$ and $\Omega_{\Lambda^*N}$, describing $\bar{K}N$
and $N(\bar{K}N)$ correlations, respectively, and chiral Lagrangians
with derivative meson--baryon couplings invariant under chiral
$SU(3)\times SU(3)$ symmetry, used for the analysis of low--energy
strong interactions \cite{ECL1,ECL2}. The frequencies
$\Omega_{\Lambda^*}$ and $\Omega_{\Lambda^*N}$ take also into account
their different strengths in contrast to \cite{KNC09}. Since the
angular momenta of the KNCs ${^1_{\bar{K}}}{\rm H}$ and
${^2_{\bar{K}}}{\rm H}$ are equal to zero, their states are defined by
the vibrational degrees of freedom, which are described by the trial
harmonic oscillator wave functions \cite{LL07}. The use of the trial
harmonic oscillator wave functions is also supported by the
short--range character of forces, producing deeply bound KNCs.

Recent experimental data \cite{DISTO1}, obtained from the analysis of
exclusive $K^+$--meson missing mass and $p \Lambda^0$ invariant mass
spectra of the final state in the $pp \to K^+ p \Lambda^0$ reaction at
the incident proton kinetic energy $T_p = 2.85\,{\rm GeV}$, have shown
a signature for the creation of a compact object $X$ with a probable
structure $X = \bar{K}NN$, decaying into the $p \Lambda^0$ state, $X
\to p + \Lambda^0$ \cite{DISTO1}.  According to \cite{DISTO1}, such a
compact object possesses an unexpected large \cite{KNC09}--\cite{KNC5}
binding energy $B_X = 103(6)\,{\rm MeV}$ and the width $\Gamma_X =
118(13)\,{\rm MeV}$ in the two--body reaction channel $pp \to K^+ X$
at high momentum transfer. For a deeply bound state $X$ with a
structure $\bar{K}NN$ and the binding energy $B_X = 103(6)\,{\rm MeV}$
the three--body pionic decay modes are suppressed \cite{DISTO1}. The
theoretical analysis of deeply bound states of this kind cannot be
carried out within the coupled--channel Faddeev equation approach
\cite{KNC1,KNC2} and the $SU(3)$ coupled--channel approach with chiral
dynamics \cite{KNC4,KNC5}. These approaches are unable to describe the
two--body reaction channels, enhanced at high momentum transfer, and
the two--body non--pionic decay modes of the deeply bound $\bar{K}NN$
state.

In this letter we propose a modified molecule model of KNCs, going
beyond the tree--approximation, and apply it to the description of the
deeply bound $\bar{K}NN$ state, observed by the DISTO Collaboration
\cite{DISTO1}. In this model we treat this $\bar{K}NN$ state as the
KNC ${^2_{\bar{K}}}{\rm H}$ with the structure $N \otimes
(\bar{K}N)_{I = 0}$. As has been pointed out in \cite{KNC09}, the
deeply bound state ${^2_{\bar{K}}}{\rm H}$ should be analysed together
with a simpler deeply bound state ${^1_{\bar{K}}}{\rm H}$, having a
structure $(\bar{K}N)_{I = 0}$, which we identify with the $\Lambda^*$
resonance with mass $m_{\Lambda^*} = 1405\,{\rm MeV}$ \cite{KNC09}
(see also \cite{COSY}--\cite{L1405}). The important role of the
$\Lambda^*$ resonance in the formation of the $X = \bar{K}NN$
resonance in the $pp \to X K^+$ reaction has been investigated
recently in \cite{DISTO2} for the kinetic energies $T_p = 2.50\,{\rm
GeV}$ and $T_p = 2.85\,{\rm GeV}$ of the incident protons. It has been
shown that the $\Lambda^*$ resonance is the doorway state for the
formation of the $\bar{K}NN$ state with a high $p\Lambda^*$ sticking
probability at $T_p = 2.85\,{\rm GeV}$.

Since the widths of the KNC ${^1_{\bar{K}}}{\rm H}$ and
${^2_{\bar{K}}}{\rm H}$ have been calculated in \cite{KNC09}, the
analytical expressions of which we use in this letter, we shall focus
on the calculation of the binding energies and other properties of
these states, such as densities.  sizes and formation probabilities in
the $pp$ reaction by taking into account all relevant interaction
channels.

\subsection{2. Analytical expression and numerical value of the binding 
energy of KNC ${^1_{\bar{K}}}{\rm H}$}

Since in \cite{KNC09} we have been restricted by the
tree--approximation, the binding energy of the KNC ${^1_{\bar{K}}}{\rm
H}$ has been determined by the Weinberg--Tomozawa (WT) $(\bar{K}N)_{I
= 0} \to (\bar{K}N)_{I = 0}$ interaction only.  Such a calculation
does not take into account the contributions of other possible
WT--interactions, making an important influence on the dynamics of the
KNC ${^1_{\bar{K}}}{\rm H}$.  Here we calculate the contribution of
all interaction channels $(\bar{K}N)_{I = 0} \to (\bar{K}N)_{I = 0}$,
$(\bar{K}N)_{I = 0} \to (\Sigma\pi)_{I = 0}$ and $(\bar{K}N)_{I = 0}
\to \Lambda\eta$ in terms of the self--energy loop corrections, where
the vertices of Feynman diagrams are defined by the
WT--interactions. These contributions are required also by a
self--consistency of the analysis of the binding energy of the KNC
${^1_{\bar{K}}}{\rm H}$ with its decay modes.  Indeed, the real parts
of these diagrams give the contributions to the mass of the KNC
${^1_{\bar{K}}}{\rm H}$, whereas the imaginary parts define the
partial widths of the kinematically allowed decay channels.

Thus, in our approach the mass of the KNC ${^1_{\bar{K}}}{\rm H}$ is
determined as
\begin{eqnarray}\label{label1}
\hspace{-0.3in}M_{{^1_{\bar{K}}}{\rm H}} = m_N + m_K -
B^{WT}_{{^1_{\bar{K}}}{\rm H}}+ \sum_X \delta
M^{(X)}_{{^1_{\bar{K}}}{\rm H}}
\end{eqnarray}
where $B^{WT}_{{^1_{\bar{K}}}{\rm H}}$ defines the tree--level
contribution, caused by the $(\bar{K}N)_{I = 0} \to (\bar{K}N)_{I =
0}$ WT--interaction. The mass--corrections $\delta
M^{(X)}_{{^1_{\bar{K}}}{\rm H}}$ are calculated for the complete set
of two--body intermediate states $X = (\bar{K}N)_{I = 0},
\Lambda^0\eta$ and $(\Sigma\pi)_{I = 0}$.  The calculation shows that
only the intermediate state $\Lambda^0\eta$ gives a finite
contribution. Its Feynman diagram is shown in Fig.\,1.
\begin{figure}
\centering \includegraphics[height=0.10\textheight]{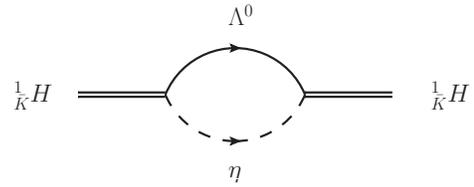}
\caption{The Feynman diagram of the main correction to the mass of the
KNC ${^1_{\bar{K}}}{\rm H}$. }
\end{figure}
The binding energy of the KNC ${^1_{\bar{K}}}{\rm H}$ with the
mass--correction is
\begin{eqnarray}\label{label2}
B_{{^1_{\bar{K}}}{\rm H}} = B^{WT}_{{^1_{\bar{K}}}{\rm H}} -
\frac{27}{512\pi}\,\frac{m_K}{F^4_{\pi}}\,\sqrt{m^2_{\eta} -
m^2_K}\Big(\frac{\mu_ {\Lambda^*}\Omega_{\Lambda^*}}{\pi}\Big)^{3/2}.
\end{eqnarray}
Setting the binding energy $B_{{^1_{\bar{K}}}{\rm H}}$ of the KNC
${^1_{\bar{K}}}{\rm H}$ equal to $B_{{^1_{\bar{K}}}{\rm H}} = m_N +
m_K - m_{\Lambda^*} = 29\,{\rm MeV}$, we get
$\Omega_{\Lambda^*} = 59.5\,{\rm MeV}$. It is larger compared with the
frequency $\Omega_{\Lambda^*} = 46.3\,{\rm MeV}$, derived in
\cite{KNC09}, thus leading also to a stronger binding of the KNC
${^2_{\bar{K}}}{\rm H}$.

Using the results, obtained in \cite{KNC09}, we calculate the width
$\Gamma_{{^1_{\!\bar{K}}}{\rm H}} = 36.4\,{\rm MeV}$ of the KNC
${^1_{\bar{K}}}{\rm H}$, which is defined by the $(\bar{K}N)_{I = 0}
\to (\Sigma\pi)_{I = 0}$ WT--interaction only. It compares well with
$\Gamma_{{^1_{\bar{K}}}{\rm H}} = 40\,{\rm MeV}$, accepted for the
description of the $\Lambda^*$ state in \cite{Akaishi1}.

\subsection{3. Analytical expression of the binding energy
 of KNC ${^2_{\bar{K}}}{\rm H}$}

The contributions to the mass of the KNC ${^2_{\bar{K}}}{\rm H}$ are
given by the Feynman diagrams in Fig.\,2.  For the binding energy of
the KNC ${^2_{\!\bar{K}}}{\rm H}$ we obtain the following expression,
including the contributions of the tree--level approximation and
self--energy loop corrections
\begin{eqnarray}\label{label3}
\hspace{-0.3in}&&B^{j'j}_{{^2_{\bar{K}}}{\rm H}} =
(B^{WT}_{{^2_{\bar{K}}}{\rm H}})^{jj'} - \sum_X\delta
M^{(X)j'j}_{{^2_{\bar{K}}}{\rm H}} = (B^{WT}_{{^2_{\bar{K}}}{\rm
H}})^{jj'} \nonumber\\
\hspace{-0.3in}&&+
\,\delta^{j'j}\,\frac{\xi}{\sqrt{2}}\,\frac{81}{8192\pi^2}
\,\frac{(m_{\eta} + m_K)^2}{F^4_{\pi}}\frac{m^2_{\eta}}{m_K} \,
(m^2_{\eta} - m^2_K)\nonumber\\
\hspace{-0.3in}&&\times\,
\Big(\frac{\mu_{\Lambda^*}\Omega_{\Lambda^*}}
{\mu_{\Lambda^*N}\Omega_{\Lambda^*N}}\Big)^{3/2} +
\,\delta^{j'j}\,\frac{1}{\sqrt{2}}\,\frac{141}{8192\pi^2}
\,\frac{(m_{\pi} + m_K)^2}{F^4_{\pi}}\nonumber\\
\hspace{-0.3in}&&\frac{m^2_{\pi}}{m_K}\,
(m^2_{\pi} - m^2_K)\,
\Big(\frac{\mu_{\Lambda^*}\Omega_{\Lambda^*}}
{\mu_{\Lambda^*N}\Omega_{\Lambda^*N}}\Big)^{3/2}.
\end{eqnarray}
Here $j$ and $j'$ are nucleon isospin indices, $\Omega_{\Lambda^*N}$
is the frequency of the relative motion of the nucleon $N$ and the
pair $\bar{K}N$, $\mu_{\Lambda^*}$ and $\mu_{\Lambda^*N}$ are the
reduced masses of the $(\bar{K}N$ and $N(\bar{K}N)$ systems,
respectively, and $\xi = 4m_N/(m_{\eta} + m_K + 4m_N)$.

The binding energy $(B^{WT}_{{^2_{\bar{K}}}{\rm H}})^{jj'}$ is
calculated in the tree--approximation and determined by the
$(\bar{K}N)_{I = 0} \to (\bar{K}N)_{I = 0}$ and $(\bar{K}N)_{I = 1}
\to (\bar{K}N)_{I = 1}$ WT--interactions. The contribution of the
WT--interaction with isospin $I = 1$ is caused by the $NN$ exchange
interaction.  The second and the third terms in Eq.(\ref{label3}) are
the mass--corrections, induced by the effective $NN\bar{K} \to
N\Lambda^0\eta, N (\Sigma \pi)_{I = 0}, N (\Sigma \pi)_{I = 1}$ and
$NN\bar{K} \to N \Lambda^0 \pi$ interactions, respectively, which
correspond to the intermediate $X = N\Lambda^0\eta, N (\Sigma \pi)_{I
= 0}, N (\Sigma \pi)_{I = 1}$ and $N \Lambda^0 \pi$ states in the
${^2_{\bar{K}}}{\rm H} \to X \to {^2_{\bar{K}}}{\rm H}$
transition. The coupling constants of these interactions are
calculated by using chiral Lagrangians. The contributions of the
$NN\bar{K} \to N (\Sigma\pi)_{I = 1}$ and $NN\bar{K} \to N \Lambda^0
\pi$ interactions appear due to the $NN$ exchange interaction.
\begin{figure} \centering
\includegraphics[height=0.10\textheight]{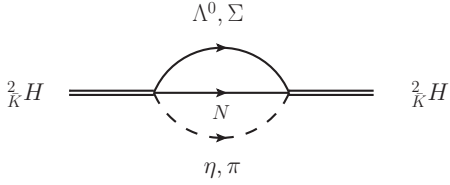}
\caption{The Feynman diagrams of the main corrections to the mass of
the KNC ${^2_{\bar{K}}}{\rm H}$. }
\end{figure}  

\subsection{4. Numerical values of binding energy, width and density of KNC 
${^2_{\bar{K}}}{\rm H}$}

For the large binding energy of the KNC ${^2_{\bar{K}}}{\rm H}$,
measured by the DISTO Collaboration \cite{DISTO1}, the pionic decay
modes $N\Sigma\pi$ are suppressed due to a very small phase volume of
the final state.  As a result the non--pionic decay modes
${^2_{\bar{K}}}{\rm H} \to N\Lambda^0$ and ${^2_{\bar{K}}}{\rm H} \to
N\Sigma$ are allowed only. The partial widths of the non--pionic decay
modes have been calculated in \cite{KNC09}. Using these results,
Eq.(\ref{label3}), and the frequency $\Omega_{\Lambda^*} = 59.5\,{\rm
MeV}$, fixed from the binding energy of the KNC ${^1_{\bar{K}}}{\rm
H}$, we obtain the binding energy $B_{{^2_{\bar{K}}}{\rm H}} =
(74.5)_{WT} + (+ 54.5)_{N\Lambda^0\eta}+(- 12.0)_{N(\Sigma +
\Lambda^0)\pi} = 118\,{\rm MeV}$ and the width
$\Gamma_{{^2_{\bar{K}}}{\rm H}} = (97)_{N\Lambda^0} + (45)_{N\Sigma} =
142\,{\rm MeV}$. They are calculated for the frequency
$\Omega_{\Lambda^*N} = 105.0\,{\rm MeV}$, providing an optimal ratio
$B_{{^2_{\bar{K}}}{\rm H}}/\Gamma_{{^2_{\bar{K}}}{\rm H}}= 0.83$ in
comparison with the experimental one $B^{(\exp)}_{{^2_{\bar{K}}}{\rm
H}}/\Gamma^{(\exp)}_{{^2_{\bar{K}}}{\rm H}}= 0.87(11)$ \cite{DISTO1}.

In addition these frequencies define the high density of the KNC
${^2_{\bar{K}}}{\rm H}$ equal to
\begin{eqnarray}\label{label4}
\hspace{-0.3in}n_{{^2_{\bar{K}}}{\rm H}}(0) =
\frac{1}{\pi^{3/2}}\,\frac{(\mu_{\Lambda^*} +
\mu_{\Lambda^*N})^3}{\displaystyle
\Big(\frac{\mu_{\Lambda^*}}{\Omega_{\Lambda^*}} +
\frac{\mu_{\Lambda^*N}}{\Omega_{\Lambda^*N}}\Big)^{3/2}} = 2.71\,n_0,
\end{eqnarray}
which agrees well with the density $n_{\bar{K}NN}(0) \sim 3\,n_0$,
estimated in \cite{DISTO1}, where $n_0 = 0.17\,{\rm fm^{-3}}$ is the
normal nuclear density.  The root mean square (rms) radius of the KNC
${^2_{\bar{K}}}{\rm H}$ is $R_{{^2_{\bar{K}}}{\rm H}}= 0.89\,{\rm fm}$
and thus is direct evidence for the compactness of the deeply bound
KNC ${^2_{\bar{K}}}{\rm H}$. As we show below the value of the rms
$R_{{^2_{\bar{K}}}{\rm H}}= 0.89\,{\rm fm}$ agrees well with the high
$\Lambda^*p$ sticking probability, observed in the $pp$ reaction at
$2.85\,{\rm GeV}$ kinetic energy of the incident protons
\cite{DISTO2}. For the KNC ${^1_{\bar{K}}}{\rm H}$ we get a low
density $n_{{^1_{\bar{K}}}{\rm H}}(0) =
(\mu_{\Lambda^*}\Omega_{\Lambda^*}/\pi)^{3/2} = 0.37\,n_0$ and a large
rms radius $R_{{^1_{\bar{K}}}{\rm H}}= 1.74\,{\rm fm}$.

The results, obtained in the molecule model, are summarised in Table I
and compared with the experimental data by the DISTO Collaboration,
which are well reproduced. In addition the widths of all partial decay
modes ${^2_{\bar{K}}}{\rm H} \to N\Lambda^0, N\Sigma^+$ and
$N\Sigma^0$ are predicted.

\begin{table}[h]
\begin{tabular}{|c|c|c|}
\hline & Molecule model & DISTO \\ \hline $(\Omega_{\Lambda^*},
\Omega_{\Lambda^*N})$& $(59.5, 105.0)$ & \vspace{0.01in}\\\hline
$B_{{^1_{\bar{K}}}{\rm H}} $ & $29.0$ & \vspace{0.01in}\\ \hline
$\Gamma_{{^1_{\bar{K}}}{\rm H}}$ & $36.4$ &\vspace{0.01in}\\\hline
$n_{{^1_{\bar{K}}}{\rm H}}(0)$ & $0.37 n_0$ &\vspace{0.01in}\\\hline
$B_{{^2_{\bar{K}}}{\rm H}}$ & $118$ & 103(6)\vspace{0.01in}\\ \hline
$\Gamma({^2_{\bar{K}}}{\rm H} \to N\Lambda^0)$ & $97$ &
\vspace{0.01in}\\ \hline $\Gamma({^2_{\bar{K}}}{\rm H} \to N\Sigma^0)$
& $15$ & \vspace{0.01in}\\ \hline $\Gamma({^2_{\bar{K}}}{\rm H} \to
N\Sigma^+)$ & $30$ &\vspace{0.01in}\\\hline
$\Gamma_{{^2_{\bar{K}}}{\rm H}}$ & $142$ & $
118(13)$\vspace{0.01in}\\\hline $n_{{^2_{\bar{K}}}{\rm H}}(0)$ & $2.71
n_0$ & $\sim 3\,n_0$\vspace{0.01in}\\\hline
\end{tabular} 
\caption{The binding energies and widths of the KNC
${^1_{\bar{K}}}{\rm H}$ and ${^2_{\bar{K}}}{\rm H}$, measured in MeV,
and their densities in terms of the normal nuclear density $n_0 =
0.17\,{\rm fm^{-3}}$. }
\end{table}

\subsection{5. Concluding discussion}

The large total width of the KNC ${^2_{\bar{K}}}{\rm H}$, such as
calculated above and observed in \cite{DISTO1}, is a strong indication
of the compactness of the $\bar{K}NN$ system as was pointed out
already in \cite{DISTO1}. Thus, one can assert that a large width
$\Gamma_{{^2_{\bar{K}}}{\rm H}} = 142\,{\rm MeV}$ of the KNC
${^2_{\bar{K}}}{\rm H}$ is fully caused by its large density
$n_{{^2_{\bar{K}}}{\rm H}}(0) = 2.71\,n_0$ \cite{DISTO1}.  In turn,
the width $\Gamma_{{^1_{\bar{K}}}{\rm H}} = 36.4\,{\rm MeV}$ of the
KNC ${^1_{\bar{K}}}{\rm H}$, which is smaller compared with the width
of the KNC ${^2_{\bar{K}}}{\rm H}$, should imply a smaller
density. This agrees qualitatively with the value
$n_{{^1_{\bar{K}}}{\rm H}}(0) = 0.37\,n_0$.

The main contributions to the binding energy of the KNC
${^2_{\bar{K}}}{\rm H}$ come from the $(\bar{K}N)_{I= 0} \to
(\bar{K}N)_{I = 0}$ and $(\bar{K}N)_{I = 1} \to (\bar{K}N)_{I = 1}$
WT--interactions and the effective $NN\bar{K} \to N \Lambda^0\eta$
interaction, the coupling constant of which has the same strength as
the $(\bar{K}N)_{I = 0} \to (\bar{K}N)_{I = 0}$ WT--interaction.

For a confirmation of the compactness of the KNC ${^2_{\bar{K}}}{\rm
H}$ in the molecule model of KNCs we can estimate the depth
$U_{\Lambda^*N}$ of the potential of the $\Lambda^*N$ interaction in
the KNC ${^2_{\bar{K}}}{\rm H}$.  We get $U_{\Lambda^*N} = -
B_{{^2_{\bar{K}}}{\rm H}} - \frac{3}{2}\, \Omega_{\Lambda^*N} = -
276\,{\rm MeV}$, which is larger than the value $U_{N(\bar{K}N)} = -
200\,{\rm MeV}$, originally proposed by Yamazaki and Akaishi
\cite{Yamazaki3}. This confirms the existence of the KNC
${^2_{\bar{K}}}{\rm H}$ as a very compact $\bar{K}NN$ system, caused
by a very strong $\Lambda^*N$ interaction.

According to \cite{Yamazaki3} (see also \cite{DISTO1}), the KNC
${^2_{\bar{K}}}{\rm H}$ should appear as a result of a sticking of the
$\Lambda^*p$ pair into the KNC ${^2_{\bar{K}}}{\rm H}$, produced in
the $pp \to K^+ \Lambda^*p \to K^+ \Lambda^0 p $ reaction. A high
$\Lambda^*p$ sticking probability has been recently observed in
\cite{DISTO2}. In the molecule model we can estimate the sticking
probability of the $\Lambda^*N$ pair in the center of mass frame as
the probability of finding the $\Lambda^*N$ pair inside the KNC
${^2_{\bar{K}}}{\rm H}$. Thus, the sticking of the $\Lambda^*N$ pair
into the KNC ${^2_{\bar{K}}}{\rm H}$ with relative momenta $p \le q <
\infty$ is given by
\begin{eqnarray}\label{label5}
\hspace{-0.3in}&&P(\Lambda^*N \to {^2_{\bar{K}}}{\rm H})(p) = \int_{q
  \ge p} |\Phi_{\Lambda^*N}(\vec{q}\,)|^2 \frac{d^3q}{(2\pi)^3}.
\end{eqnarray}
where $\Phi_{\Lambda^*N}(\vec{q}\,)$ is the wave function of a
relative motion of the $\Lambda^*N$ pair inside the KNC
${^2_{\bar{K}}}{\rm H}$ \cite{KNC09}.  Since the $\Lambda^*N$ pair
should be confined inside the KNC ${^2_{\bar{K}}}{\rm H}$, the
confining area is restricted by a diameter of order of $D =
2R_{{^2_{\bar{K}}}{\rm H}} = 1.78\,{\rm fm}$. This defines the minimum
relative momentum $p \sim 1/D = 111\,{\rm MeV/c}$ for the $\lambda^*N$
sticking in the $pp$ reaction. The sticking probability $P(\Lambda^*N
\to {^2_{\bar{K}}}{\rm H})(p) \sim 0.94$, calculated for $p \sim 1/D =
111\,{\rm MeV/c}$, agrees well with the experimental data on the high
sticking probability of the $p\Lambda^*$ pair, observed in the
reaction $pp \to K^+p\Lambda^0$ at the kinetic energy $T_p =
2.85\,{\rm GeV}$ of the incident proton \cite{DISTO2}.

This research was partly supported by the DFG cluster of excellence
"Origin and Structure of the Universe" of the Technische Universit\"at
M\"unchen and by the Austrian ``Fonds zur F\"orderung der
Wissenschaftlichen Forschung'' (FWF) under contract P19487-N16.

\end{document}